\documentclass[fleqn,11pt]{wlscirep}
\usepackage{wrapfig}
\usepackage{graphicx}
\usepackage{amsmath}
\usepackage{textcomp}
\usepackage{siunitx}
\usepackage{sidecap}
\usepackage{float} 
\usepackage{cite}
\usepackage{caption}
\usepackage{subcaption}
\usepackage[percent]{overpic}
\usepackage{pdfpages}

\usepackage{svg}

\title{Elimination of periodic nonlinearities of actuators with internal periodic processes} 
\author[1,*]{Nils Frederik Hasselmann}
\author[1]{Maxime Nicloux}
\author[1]{Alexander Sell}

\affil[1]{Airbus Defence and Space GmbH, Advanced Projects Germany, Friedrichshafen, Germany}
\affil[*]{nils\textunderscore frederik.hasselmann@airbus.com}


\begin{abstract}
  We discuss an algorithm that eliminates the periodic non-linearity of a bimorph piezo actuator by modifying the phase of the voltage waveforms used for driving the actuator. The procedure presented allows for a separate optimization of the motion of the actuator groups in order to obtain the properties desired (such as a specific stepsize or maximum force). The resulting response of a bimorph piezo drive, also under different loads, is experimentally investigated. The optimization process can be applied to other actuators relying on internal periodic processes, if the motion is reproducible, and is limited by measurement noise or manufacturing tolerances only.
\end{abstract}

\begin{document}

\flushbottom
\maketitle
\thispagestyle{empty}
\section*{Introduction}
While classic piezo actuators, made from a stack of piezoelectric crystals, allow for an actuation precision better than $\SI{1}{\nano\metre}$, their total stroke is limited to a fraction of the stacks' dimensions.\cite{mohith_recent_2020} Bimorph piezo drives on the other hand maintain the advantage of ultra high precision positioning capabilities while being also able to perform strokes larger than the the piezo stacks themselves.\cite{den_heijer_improving_2014,silvestri_piezoelectric_2017} In case of walking leg drives, this is because a rod is clamped down on and transported by one or more sets of two actuator groups (legs), thus allowing for a stroke that is equal to the length of the rod less the separation of the tips of the outer piezo legs.\cite{uzunovic_piezo_2015,merry_modeling_2010}
However, this operating principle requires a periodic motion that also yields an undesired periodic nonlinearity of the actuated variable.\cite{merry_using_2009}  
\begin{figure}[hbt!]
     \centering
     
 \begin{subfigure}{0.3\textwidth}
    \includegraphics[width=\textwidth]{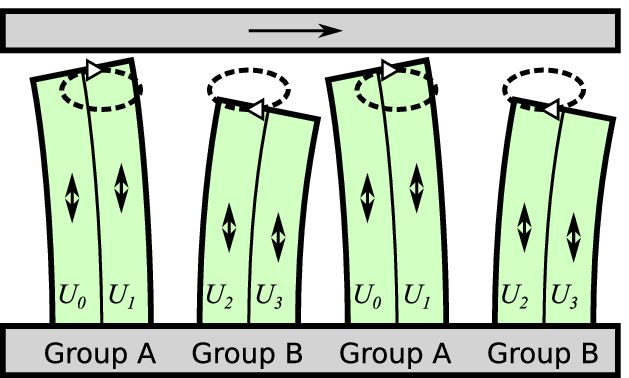}
    \caption{ }
        \label{fig:a}
    \end{subfigure} 
 \begin{subfigure}{0.3\textwidth}
    \includegraphics[width=\textwidth]{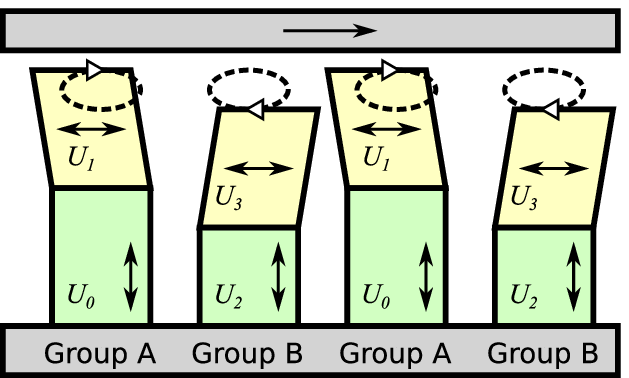}
     \caption{ }
        \label{fig:b}    
    \end{subfigure} 
  \begin{subfigure}{0.226\textwidth}
    \includegraphics[width=\textwidth]{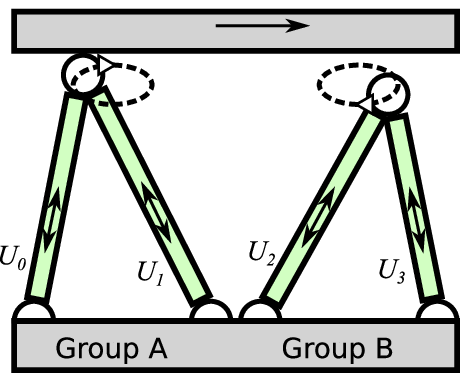}
     \caption{ }
        \label{fig:c}    
    \end{subfigure} 
        \caption{Typical geometries for piezo actuators of the bimorph type each featuring at least two groups consisting of two elements to each of which a voltage ($U_0$ to $U_3$) is applied. In a) the bending type is shown. Prominent examples are the Piezo LEGs by Piezomotor or the NEXACT by Physik Instrumente; b) shows the clamp-sheer type drive, an example would be the NEXLINE by Physik Instrumente; c) features the PICMAWalk developed by Physik Instrumente. The dashed ellipses represent example trajectories of the tip of the legs. }
        \label{fig:1}
\end{figure}
While there are several types of these piezo drives commercially available that have a different design in detail (see Fig.\ref{fig:1}), they all intrinsically suffer from periodic nonlinearities: When unoptimized, symmetric voltage waveforms are used for driving the actuator, the hand over of the rod is performed such that the legs at least partially hinder one another: The motion slows down during the hand-over.
While a model based optimization of bimorph piezo drives is limited by the accuracy of the model, the method presented here is limited only by the accuracy and repeatability of the measurement device used during optimization.\cite{szufnarowski_two-dimensional_2012}
The procedure yields in principle improvements for both open-loop and closed-loop operated systems, as in closed loop operation the control effort is reduced due to a more linear response.\cite{merry_using_2009,witvoet_2015,li_dynamic_2019}
Furthermore, it can be applied to systems with several actuated degrees of freedom as long as the drives can be separately addressed and the resulting motion can be measured; the motion needs to have a slower dynamic than the lowest resonance frequency and the movement sequence must not change for typical speeds. E.g. inertia based friction drives can not be optimized using the presented method.
\section*{Methods}
\subsection*{Principle of the optimization procedure}
The input of an actuator, whose response is to be linearized, shall be a phase $\phi$ that takes into consideration the internal periodicity of the actuator. A phase value of $2\pi$ thus corresponds to one internal cycle (for example one revolution of a spindle or one step of a walking motion).
The procedure presented here requires, that the actuated variable $a$ (position, angle, etc.) can be measured as a function of that phase.\newline
The general working principle of the method is as follows: The measured actuator position $a\left(\phi\right)$ is filtered such that a function $A\left(\phi\right)$ containing only the linear increase with $\phi$ (with a slope of $\frac{c_0}{2\pi}$) as well as harmonics of the fundamental frequency is left. By recording many actuator periods (typically more than 16) and filtering of harmonics, the procedure becomes thus resistant to measurement noise, but also requires that the measurement device used  (capacitive sensors, interferometers or optical encoders) during the initial optimization process has a much smaller (periodic) nonlinearity than the actuator to be optimized; or a different periodicity.
By solving the equation $A\left(\psi\right)=\frac{c_0}{2\pi}\phi$, a function $\psi\left(\phi\right)$ that is periodic except for a linear slope can be determined. Using this function as input signal of the actuator instead, the actuator position now follows by design the function $A\left(\psi\left(\phi\right)\right)=\frac{c_0}{2\pi}\phi$, that is linear in its response to the new input. Therefore, all deterministic periodic perturbations are eliminated from the resulting motion (within the limits imposed by measurement accuracy and repeatability).
In case of the internal processes suffering from hysteresis (e.g. when piezoelectric crystals are driven by voltage amplifiers), the procedure can be applied in an iterative fashion $\psi_1\left(\psi_2\left(...\left(\psi_n\left(\phi\right)\right)\right)\right)$. Details on the practical implementation are given in the appendix.

\subsection*{Optimizing initial waveforms to yield a more linear response}
In order to obtain a linear response of a piezo friction drive,
one should use a waveform that is derived using a model-based approach as a starting point for the optimization procedure discussed before.\cite{merry_modeling_2011,szufnarowski_two-dimensional_2012}
In the simplest form (without hysteresis effects, for a beam of high aspect ratio and small deflections compared to the beams' dimensions), one can use a linear model based on Euler-Bernoulli beam theory in order to describe the two dimensional trajectory of the tip of a single leg depending on the voltages applied to a leg ($x,y$-coordinates as defined by convention): 
\begin{equation}\label{eqn:1}
\vec{D}\left(\phi\right)=
 \begin{pmatrix}
    x_A\left(\phi\right)\\
    y_A\left(\phi\right)
 \end{pmatrix}=
 \begin{pmatrix}
    m_{11}&m_{12}\\
    m_{21}&m_{22}
 \end{pmatrix}
 \begin{pmatrix}
    U_0\left(\phi\right)\\
    U_1\left(\phi\right)
\end{pmatrix}=
\boldsymbol{M}\cdot\vec{U}
\end{equation}
For the second group of legs 'B', the phase is shifted by $\pi$, assuming identical properties of that group.
Different actuator geometries (see Fig.\ref{fig:1} ) can be taken into account by choosing the coefficients of the matrix $\boldsymbol{M}$ accordingly: For walking leg piezo actuators with bending legs $m_{12}=-m_{11}$ and $m_{22}=m_{21}$, while for clamp-sheer actuators $m_{11}=m_{22}=0$. Therefore, only 2 coefficients have to be determined for actuators of these geometries from the measurements. A trajectory yielding a linear motion should fulfill following conditions (also see references \cite{johansson_fine_2004,merry_using_2009}):
\begin{itemize}
    \item $x_A\left(\phi\right)$ and $y_A\left(\phi\right)$ are $2\pi$-periodic
    \item There is exactly one phase $\phi=\phi_0$ in the interval $[0,2\pi[$, for which $y_A\left(\phi_0\right)=y_A\left(\phi_0+\pi\right)$ with $y_A\left(\phi_0\right)<y_A\left(\phi\right)$ for $\phi\in ]\phi_0,\phi_0+\pi[$ as well as $y_A\left(\phi_0\right)>y_A\left(\phi\right)$ for $\phi\in ]\phi_0-\pi,\phi_0[$. For $\phi\in ]\phi_0,\phi_0+\pi[$, actuator group A is thus responsible for transporting the rod (also see Fig.\ref{eqn:1}).
  \item The lateral speed of the load bearing actuator shall be as constant as possible in order to yield a steady motion of the actuator, especially in the hand-over region:$\frac{d x_A}{d\phi}=\textrm{const.}$ for $\phi\in [\phi_0,\phi_0+\pi]$ 
  \item The movements in both axis shall be as smooth as possible; at least continuously differentiable.
   \item The absolute vertical velocity of the load bearing actuator group should be smaller than that of the non-load bearing group, and as small as possible at handover points.
  \end{itemize}
  Two sets ($i=1,2$) of simple piece wise defined functions fulfilling these  requirements are:
  \begin{align*}
      x_{A_i}\left(\phi\right)=\frac{3\sqrt{6}}{4\pi}x_0
    \begin{cases}
        \phi & \text{, } \phi \in \left[-\frac{\pi}{2},\frac{\pi}{2}\right]\\
        \frac{4}{\pi^2}\left(\phi-\pi\right)^3-2\left(\phi-\pi\right) & \text{, } \phi \in \left]\frac{\pi}{2},\frac{3\pi}{2}\right[
    \end{cases}\mathrm{\quad and \quad}
      y_{A_i}\left(\phi\right)=\frac{1}{2}\left(1+\cos{\left(\frac{\phi^2}{\pi}\right)}\right)^i \text{, } \phi \in \left]-\pi,\pi\right]
  \end{align*}
  A value of $i=1$ will yield a rounded triangular shape as trajectory (waveform 2) and $i=2$ results in a trajectory the shape of a box with rounded edges (waveform 3).
  In order to obtain the voltages yielding approximately the desired trajectory, one simply inverts the model matrix and multiplies it with the trajectory. Waveform 1 is based on trivial sine/cosine functions and corresponds to an elliptical trajectory with an offset. 
\section*{Results}
In this section, data on the validity of the piezo model, the performance of the optimization procedure as well as on the characteristics of the LT40 piezo actuator is presented.
\subsection*{Characterization of the stepsize for different loads and validation of the piezo model}
\begin{figure}[hbt!]
  \centering
   \begin{subfigure}{0.27\textwidth}
    \includegraphics[width=\textwidth]{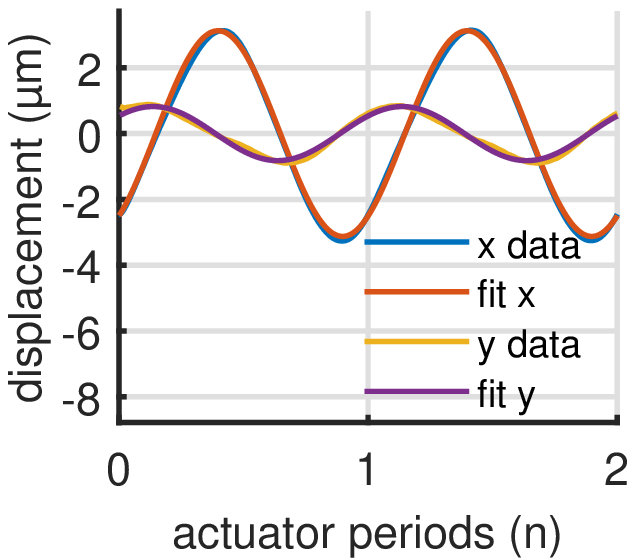}
     \caption{ }
       \label{fig:a}    
   \end{subfigure} 
     \begin{subfigure}{0.27\textwidth}
    \includegraphics[width=\textwidth]{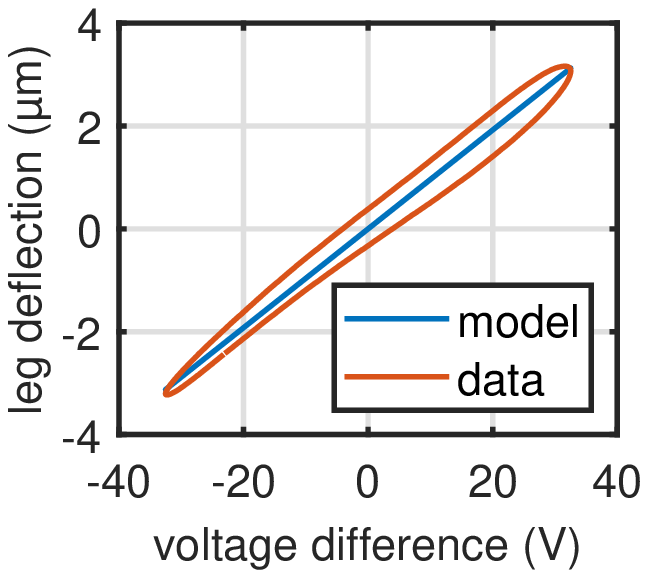}
     \caption{ }
        \label{fig:b}
    \end{subfigure} 
    \begin{subfigure}{0.27\textwidth}
    \includegraphics[width=\textwidth]{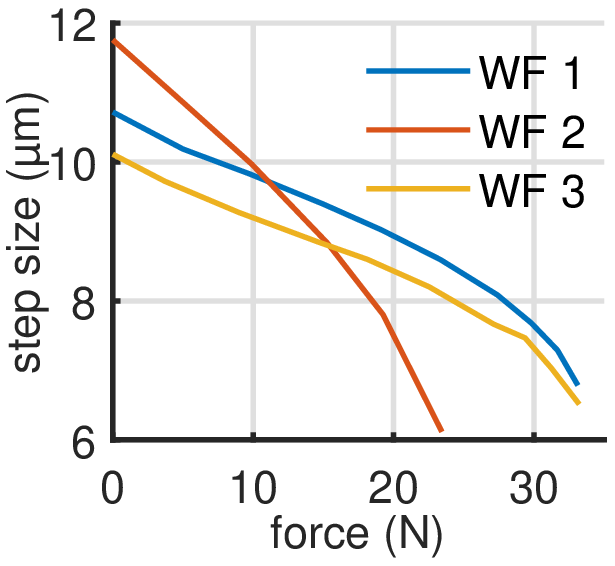}
     \caption{ }
        \label{fig:c}    
    \end{subfigure} 
   
        \caption{a) The trajectory of the tips of the legs plotted versus the actuator period; the linear model is fitted to the data in order to determine the model coefficients and verify the model. b) The presence of hysteresis is confirmed by plotting the leg deflection in x versus the difference in voltage applied to both elements of a leg. c) Experimental step size dependency on the external force for the optimized waveforms.}
        \label{fig:2}
\end{figure}

In order to validate the piezo model, the piezo actuator was rewired and driven such that the sets of legs attacking the ceramic rod from the opposite side were moving in tandem with their counterparts. Both actuator groups were performing the same elliptic motion (waveform 1), so that the rod was oscillating in x-and y-direction. The model presented in equation \ref{eqn:1} is fitted to the data with a R-square value greater than 0.98 (Fig. \ref{fig:2}a).
The model may thus generate an initial waveform to be used as starting point for the optimization procedure. Plotting the leg deflection in x-direction (travel direction) versus the difference of the voltages applied to one leg yields a linear relationship in case there are no nonlinear effects present such as hysteresis and the piezo leg is accurately described by the linear model. This is not expected, and this expectation is confirmed by the distinct hysteresis loop that the experimental data exhibits (Fig. \ref{fig:2}b). A nonlinear model would provide a better starting point. A distinct hysteresis demands, that the procedure has to be applied in an iterative manner.
In Figure \ref{fig:2}c, a comparison of the dependency of step size from load is shown for the three waveforms defined earlier when optimized. While waveform 2 allows for the largest step size, it also produces the smallest maximum force. The general behavior observed is in agreement with the literature: Increasing the load first results in a linear reduction of step size, while for higher load slippage of the legs sets in, resulting in a drastically reduced step size (stalling).\cite{silvestri_piezoelectric_2017} This results in increased wear.   
\subsection*{Scaling of waveforms}
In order to investigate the behaviour of the piezo under a variety of loads when scaling the waveforms, the step size in dependence of the actuator load was recorded for different scales and type thereof: Both x- and y-motion were varied. When scaling the x-motion, one would expect that the  difference in clamping force on the rod defined by the difference between y-position of transporting and non-load bearing legs, is practically unaffected. This assumes that both tips are always in contact with the rod at all times. Thus, stalling should set in independent of the scale, while the step size should be proportional to the scaling.
\begin{figure}[hbt!]
\centering
  \begin{subfigure}{0.48\textwidth}
    \includegraphics[width=\textwidth]{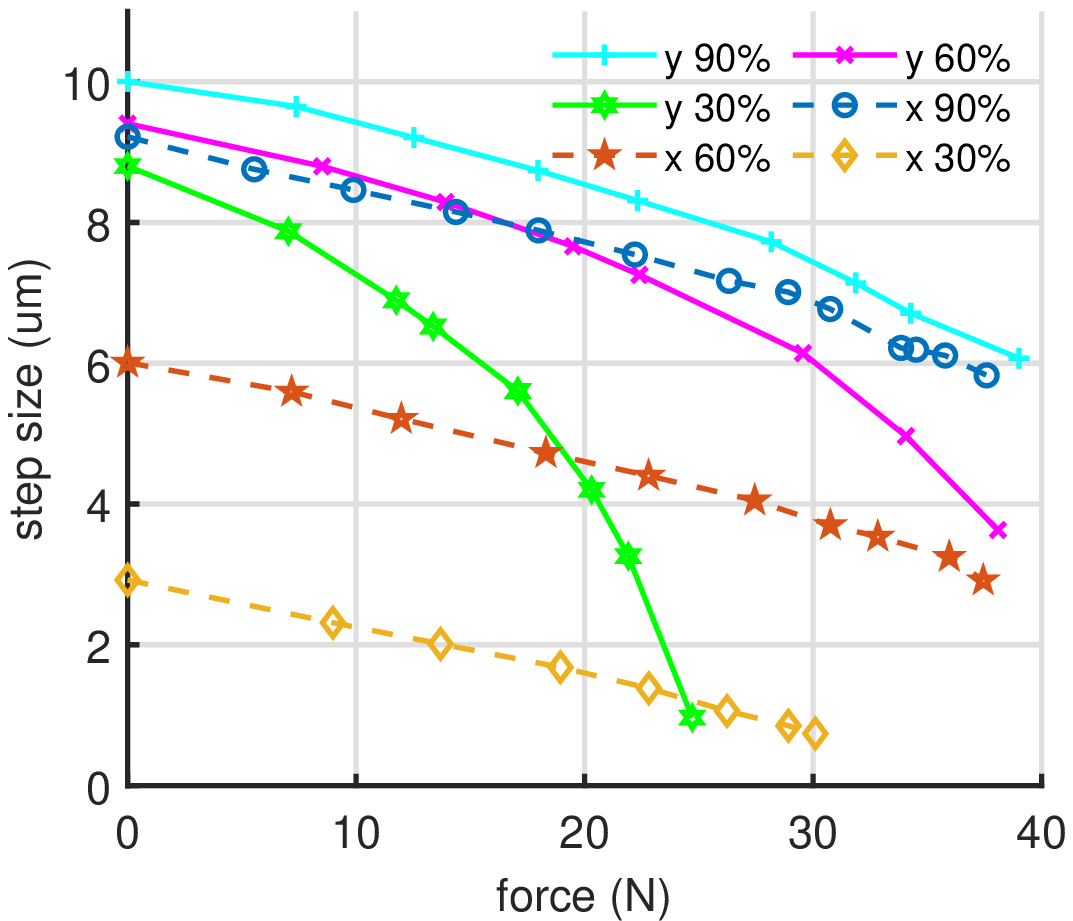}
    \caption{ }
        \label{fig:a}
    \end{subfigure}
    \begin{subfigure}{0.48\textwidth}
    \includegraphics[width=\textwidth]{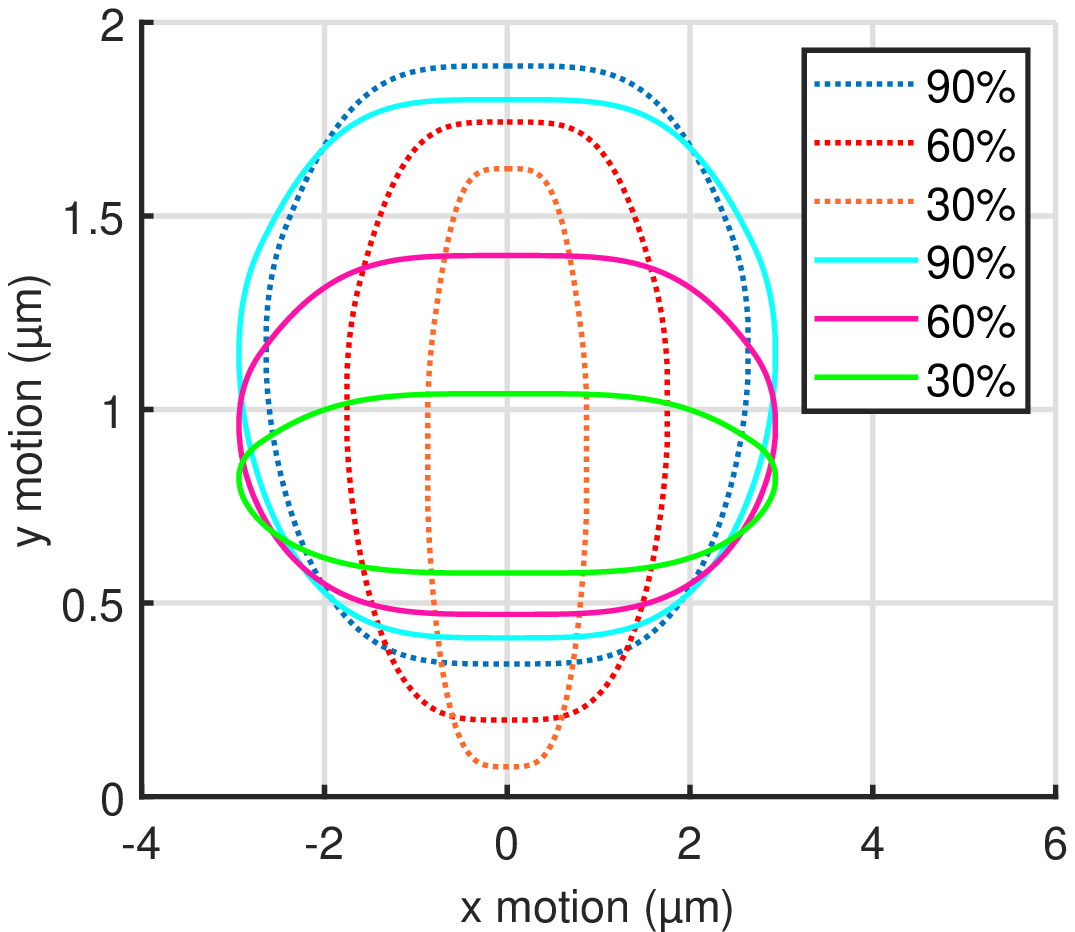}
    \caption{ }
        \label{fig:b}    
    \end{subfigure} 
        \caption{a) Step size versus external load when the trajectory of waveform 2 is scaled in x- and y-direction (dotted and full lines); b) Tip trajectories predicted by the linear model for different scales: $x$ is the lateral deflection of a leg, $y$ the elongation.}
        \label{fig:3}
\end{figure}
However, scaling the y-motion will reduce the difference in clamping forces on the rod and as a result, one would expect stalling to set in earlier compromising the maximum obtainable actuator force while also affecting the step size and increasing wear. These expectations are confirmed by the experimental data shown in Figure \ref{fig:3}. It is worth noting, that the pretension of the legs is also affected by the offset changing as a side effect of the scaling in our case. It is worth noting that the overall pretension of the piezo elements is defined by a leaf spring; as the largest possible step size was of interest and reduced pretensioning due to a sightly less deformed leaf spring has minor effects only, no action was deemed necessary. 
\subsection*{Iterative optimization of the waveforms}
When applying the optimization procedure in an iterative fashion, the termination criterion is, that the non-linear periodic error is converging due to having reached limits imposed by measurement noise or manufacturing tolerances.
\begin{figure}[hbt!]
     \centering
   \begin{subfigure}{0.48\textwidth}
    \includegraphics[width=\textwidth]{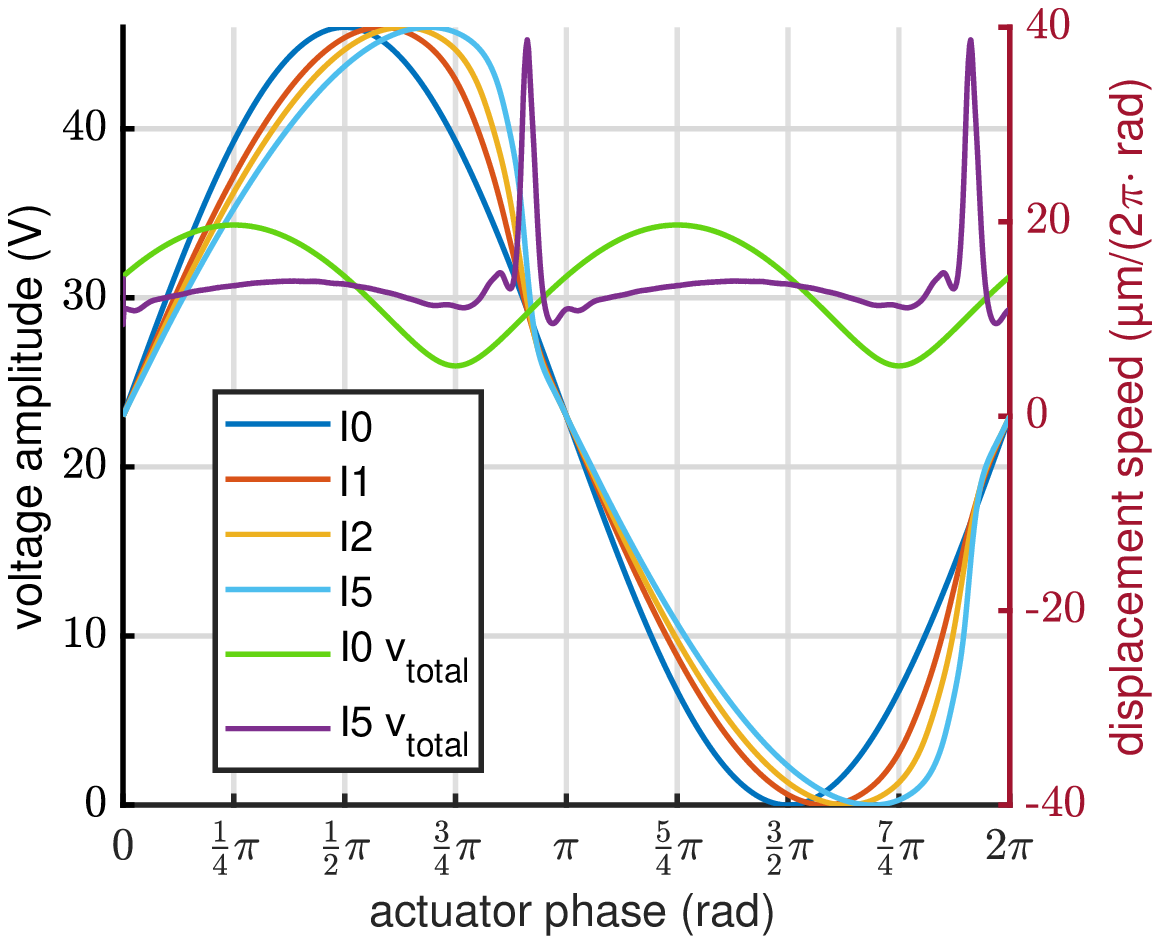}
    \caption{ }
        \label{fig:a}
    \end{subfigure}
    \begin{subfigure}{0.48\textwidth}
    \includegraphics[width=\textwidth]{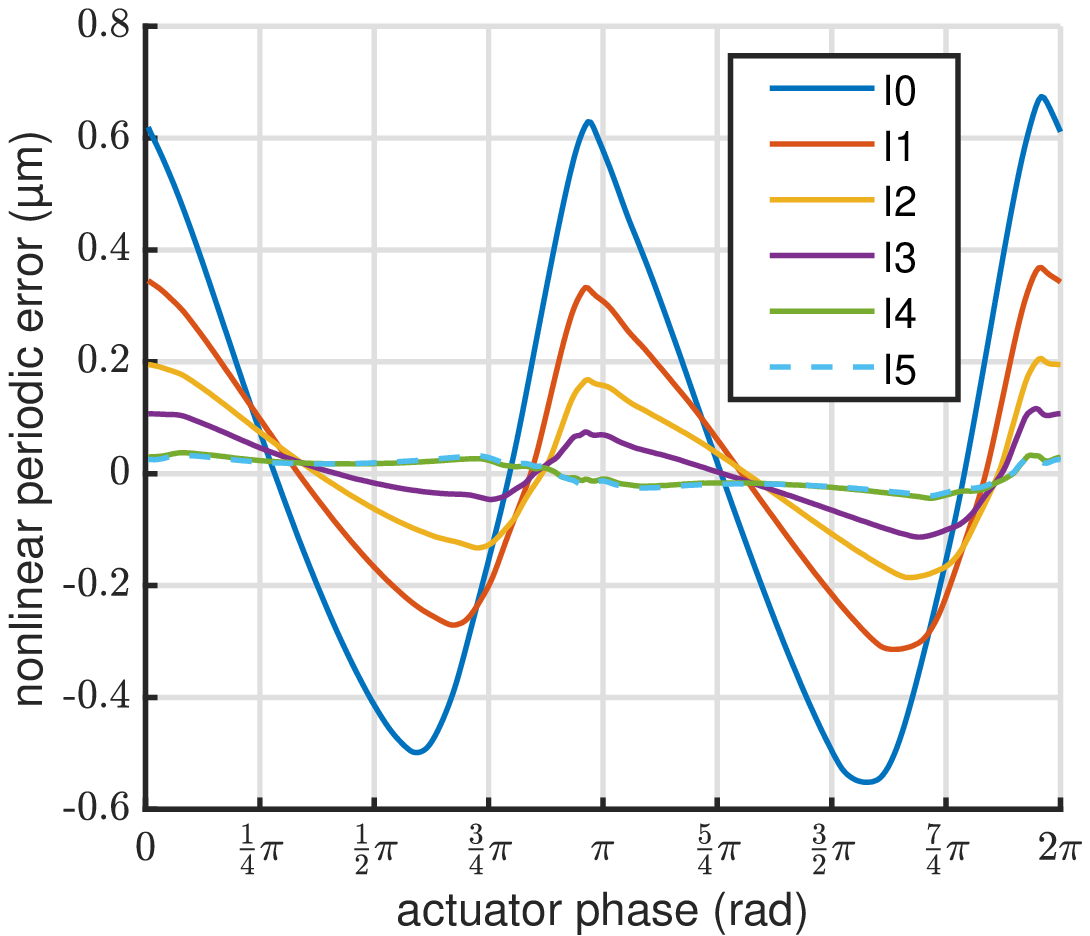}
    \caption{ }
        \label{fig:b}    
    \end{subfigure}
     \begin{subfigure}{0.48\textwidth}
    \includegraphics[width=\textwidth]{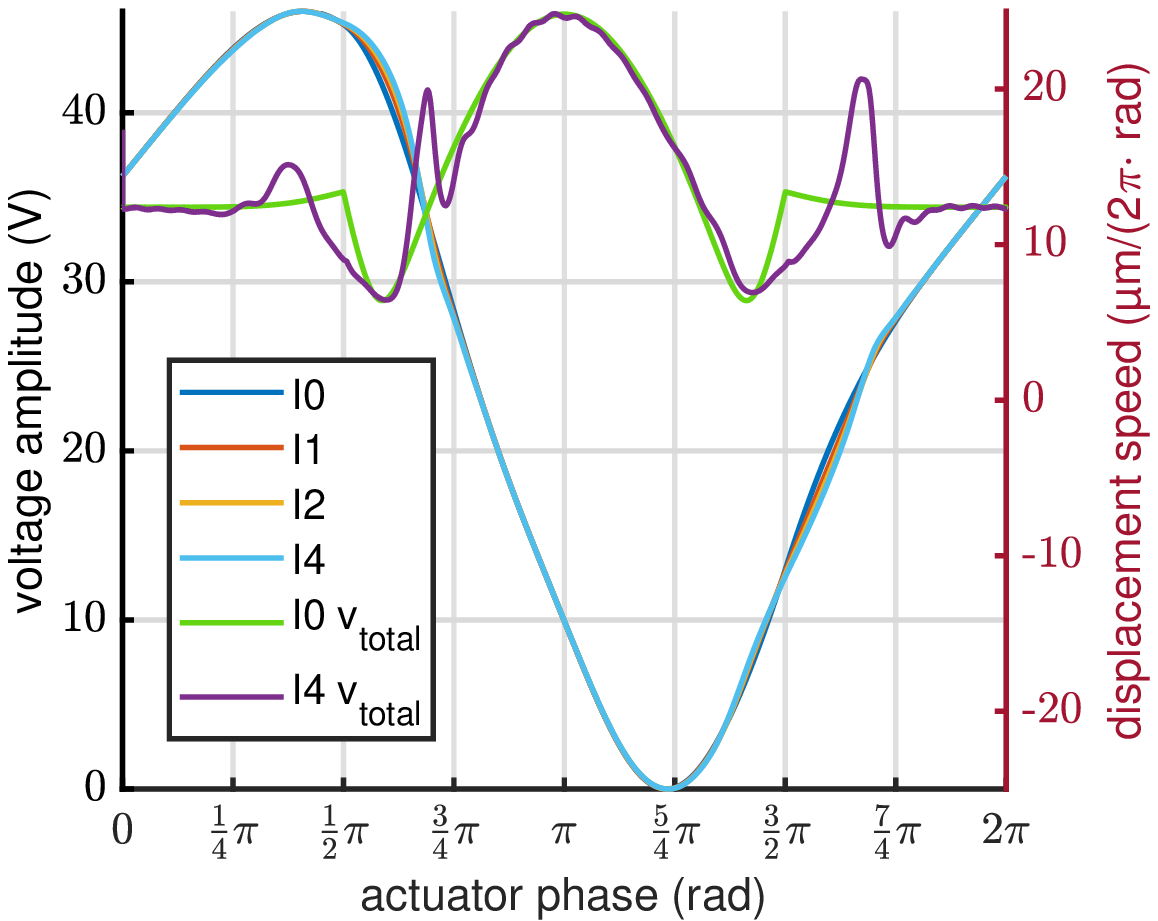}
    \caption{ }
        \label{fig:c}
    \end{subfigure}
    \begin{subfigure}{0.48\textwidth}
    \includegraphics[width=\textwidth]{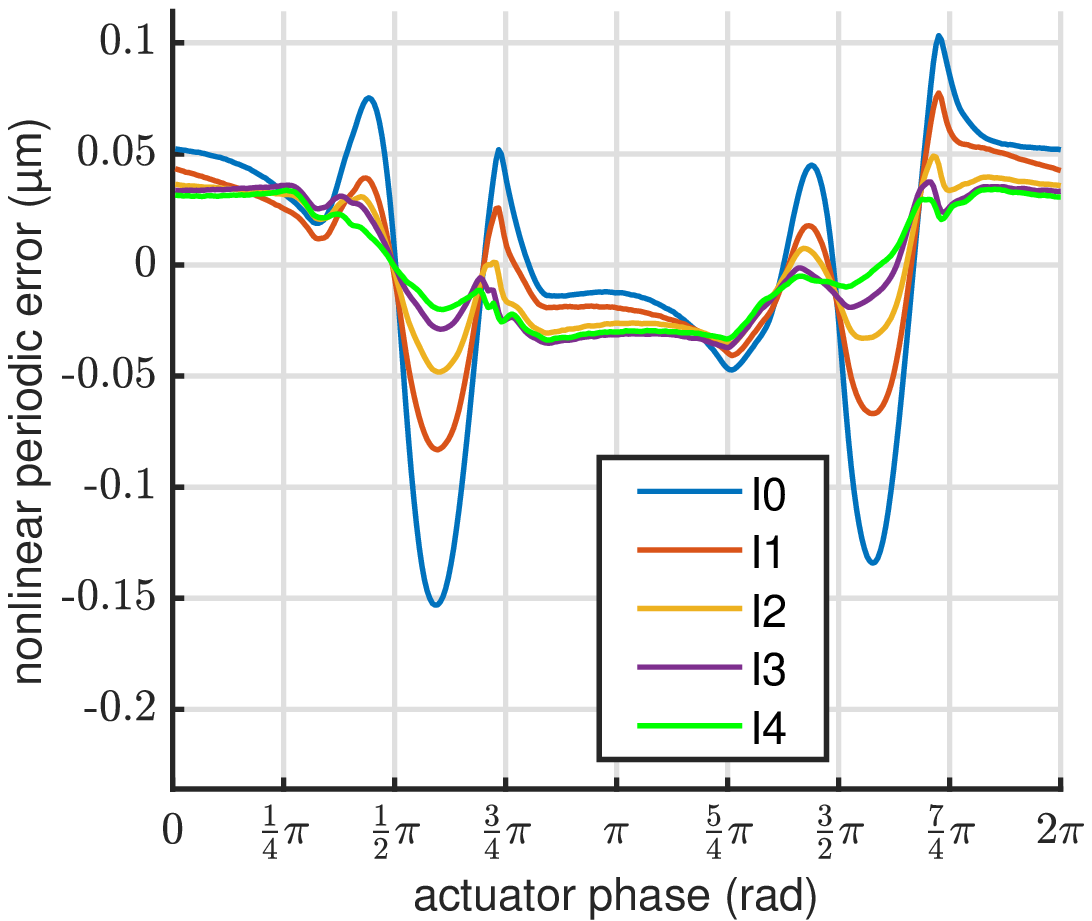}
    \caption{ }
        \label{fig:d}    
 \end{subfigure}
        \caption{a+c) Plotted is the voltage $U_0$  for waveforms 1 (a) and 3 (c) over the actuator phase for different iterations of the optimization process. Iteration 0 is the original waveform. Total speed along the trajectory of the piezo tips in units of $\si{\micro\metre}$ per actuator period for the unoptimized and the optimized case is estimated by the linear model. In b+d) the experimental data on the periodic non-linearities is shown versus the actuator period for a number of iterations.}
        \label{fig:4}
\end{figure}
In Figure \ref{fig:4}a, the resulting deformed voltage driving waveform for one of the two voltages acting on a leg is shown for the LT40 actuator. Additionally, the speed of the piezo tip along the trajectory is shown based on the linear piezo model for the unoptimized and optimized function. The optimization alters the speed to be more constant along the trajectory with spikes at the handover points, where corrections are required in order to yield a more linear motion of the ceramic rod. If one subtracts the linear trend of the motion, one can better compare the effect of the optimization on the periodic deviation or error (see Fig. \ref{fig:4}b): With each iteration, the periodic error is reduced until converging at a value about one order of magnitude below the initial one.
The proposed waveform with $i=2$ (waveform 3, Fig. \ref{fig:4}c+d) shows an analogous behavior as a result of the optimization. However, the initial periodic nonlinearity error is lower than for the trivial sine based waveform. Less iterations are required and the final waveform is not deformed as much as waveform 3 since a model-based trajectory was used as a starting point, fulfilling all but the speed requirement imposed on a trajectory that yields a highly linear response. The residual deviation is dominated by a contribution with a frequency of 1 per period, which may stem from the fact, that the different legs behave slightly differently (piezo expansion coefficients due to contacting of the electrodes, manufacturing tolerances of stack dimensions,...). So far, it was assumed that all legs behave the same and merely shifting the phase upon optimizing the waveform is sufficient in order to be limited by measurement noise only. In practice, better results may be obtained by scaling the voltages for actuator group A or B to compensate for the different effective piezo-coefficients of the other group or even optimizing them separately.
\subsection*{Changing the direction of travel using optimized waveforms}
\begin{wrapfigure}{r}{0.30\textwidth}
  \begin{center}
    \includegraphics[width=0.28\textwidth]{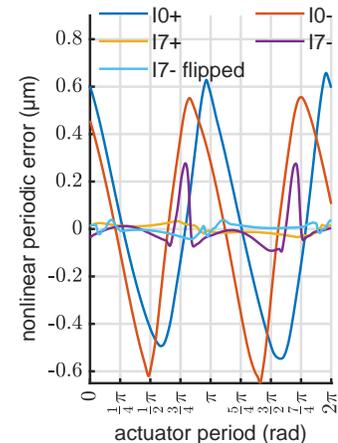}
  \end{center}
  \caption{Data on the nonlinear periodic error for travel in the nominal and the reverse direction using various waveforms, for details see text.}
   \label{fig:5}
\end{wrapfigure}

The optimization of a waveform yields a more linear response for one direction of travel only; moving in the other direction, there is still a nonlinear periodic error. In case, the piezo legs are symmetric in their response, the same waveform can be used for the reverse direction after switching the voltages $U_0$ and $U_1$ as well as $U_2$ and $U_3$ and time inverting the actuator phase $\Psi(\phi)$. The experimentally determined response for these scenarios is plotted in Figure \ref{fig:5} in terms of the deviation from a linear motion: The periodic non-linearity in the forward direction (yellow, waveform 3, optimized) is smallest.
 In reverse direction (purple, waveform 3, optimized), the performance is slightly worse, as would be expected due to symmetry considerations.
However, switching voltages and time inverting the optimized waveform 3 (light blue) yields similar results for travel in reverse direction. Doing so allows both legs to maintain their position when switching to the waveform look-up table for travel in reverse direction. Results for the unoptimized waveform in forward (dark blue) and backward direction of travel (red) are shown for comparison.
\section*{Conclusion}
A procedure for linearizing the response of actuators with internal periodic processes yielding a periodic nonlinearity was presented and experimentally validated using a bimorph piezo actuator as an example. In addition, a number of demands were imposed on the trajectory of the legs of such an actuator in order to generate voltage waveforms providing a better starting point for the optimization. Using the voltage waveforms derived, the residual motion is caused mainly by hysteresis not accounted for in the linear model and thus requires an iterative optimization. Compared to a purely model based approach, the result was initially thought to be only limited by the measurement accuracy and repeatability. The data recorded showed, that manufacturing differences of each leg have to be accounted for as well in both approaches. Nevertheless, depending on the initial waveform used, the nonlinear periodic error was reduced by up to one order of magnitude by applying the proposed optimization procedure iteratively. Furthermore, closed loop operation is much simplified as symmetries allow for usage of one optimized waveform for both directions of travel. Scaling of the trajectories of the piezo tips showed, that variations in x-direction yielded a different step size without comprimising the maximum actuator force, while scaling y only or both x and y will always result in a reduced maximum actuator force with stalling setting in earlier.
\section*{Acknowledgements}
The authors would like to thank Gerd Witvoet for fruitful discussions, Tim Vogel for supporting one of the preliminary measurements and acknowledge funding by the German BMWi via DLR grant 50OQ1302.

\bibliography{BIB}
\section*{Appendix}
\subsection*{Practical implementation:}
\begin{itemize}
    \item A time dependent phase $\phi\left(t\right)=\omega_0\cdot t$ is applied to the actuator for at least one actuator period, preferably $n\geq16\in\mathbb{N}$. The more actuator periods are recorded, the better the suppression of non-periodic noise.
    \item The actuator position is measured at equidistant, discrete times $t_j=\frac{2\pi}{m\omega_0}\cdot j=\frac{\Delta\phi}{\omega_0}\cdot j=\frac{1}{\omega_0}\cdot\phi_j$ $(m,j)\in\mathbb{N}$, with at least $m=16$ measurements being recorded per actuator period. Measurement signal is hence $a_j=\left(\phi_j\right)$.
    \item The value $b_j=a_{j+1}-a_j$ with $\left(j=0....N-1;N=n\cdot m\right)$, which is proportional to the numerical derivative at the times $t_j+\frac{\Delta t}{2}$, effectively removes the nominal linear increase with phase. This step is beneficial for the filtering, as a linear slope would result in a broad-band spectrum of harmonics.
    \item The result is filtered in order to separate periodic contributions from non-periodic noise: First the time-discrete Fourier coefficients are calculated:
    \begin{align}
        c_k=\frac{1}{n}\sum\limits_{j=0}^{N-1}b_j e^{-i\frac{2\pi}{m}j\cdot k}, k=0...\frac{m}{2}-1
    \end{align}
    After numerical integration, the filtered, strictly periodic signal $A\left(\phi\right)$ can be written as:
    \begin{align}
        A\left(\phi\right)=\frac{c_0}{2\pi}\phi+\sum\limits_{k=1}^{\frac{m}{2}-1}\Re\left(\frac{c_k}{i\pi k}\left[e^{i k\left(\phi-\frac{1}{2}\Delta\phi\right)}-e^{-\frac{1}{2}i k \Delta \phi}\right]\right)
    \end{align}
    The nominal speed of the drive with respect to the input phase is $\frac{c_0}{2\pi}$, the other coefficients correspond to the amplitudes of undesired harmonics resulting from the inner workings of the actuator. Note, that the sum is constructed so that $A\left(0\right)=0$.
    \item Now, the equation $A\left(\psi\left(\phi\right)\right)=\frac{c_0}{2\pi}\phi$ is numerically solved for $\psi\left(\phi\right)$. This is only possible in case $A\left(\psi\left(\phi\right)\right)$ is monotonous in $\psi$, which is thus a condition that has to be obeyed in order for the procedure to work.
    Solving for the desired function is trivial, as it merely entails plugging values of the interval $[0,2\pi[$ into $A\left(\psi\right)$, solving for $\phi(\psi)$ and inverting that function.
    \item If desired, the procedure can be repeated using the current function $\psi(\omega_0\cdot t)$.
   
\end{itemize}
\subsection*{Experimental setup}
The experimental setup consists of a PiezoLeg LT40 walking leg piezo actuator by PiezoMotor$^\mathrm{\textregistered}$ driven by a voltage amplifier (CEDAT Technologies LC75B power supply, two LA75B two channel amplifiers), that is connected to a linear stage, on which a mirror and a glass grating ruler are mounted. The mirrors' position is read-out using a SIOS three beam interferometer of the type SP 5000 TR and the glass rulers' position is determined using a Celera MII 6000V optical encoder with a $\SI{1.2207}{\nano\metre}$ interpolated resolution.
A FPGA of the type PCI 7833R by National Instruments is used for generating the voltage waveforms fed to the voltage amplifier as well as for processing of the encoder signals. The data is transferred to a host code running on the host PC, where it is further processed and stored. The host code also provides a graphic user interface and controls.
For force measurements, a capacitive force sensor (Kistler Type 9207) is mounted on the linear stage that then pulls the upper partner of a friction pair instead of moving freely. Weighing down the upper friction partner allows for measurement of the actuator characteristics under different loads. Readout is performed using a charge amplifier (Kistler Type 5011). The analog output is digitized using the DAC of the FPGA. The load due to the rolling resistance of the linear stage is separately measured.

\end{document}